\documentclass[preprintnumbers,amsmath,amssymb,prd,12pt,floatfix,superscriptaddress,nofootinbib]{revtex4}
\usepackage{graphicx}
\usepackage{epsfig}
\usepackage{bm}
\usepackage{amsfonts}
\def\be {\begin{equation}}
\def\ee {\end{equation}}
\def\ba {\begin{eqnarray}}
\def\ea {\end{eqnarray}}
\begin{document}

\title{Cardy-Verlinde formula for an axially symmetric dilaton-axion black hole}

\author{Mubasher Jamil}
\email{mjamil@camp.nust.edu.pk} \affiliation{Center for Advanced
Mathematics and Physics, National University of Sciences and
Technology, Rawalpindi, 46000, Pakistan}

\author{M. Akbar}
\email{makbar@camp.nust.edu.pk} \affiliation{Center for Advanced
Mathematics and Physics, National University of Sciences and
Technology, Rawalpindi, 46000, Pakistan}

\author{M. R. Setare}
\email{rezakord@ipm.ir} \affiliation{Department of Science, Payame
Noor University, Bijar, Iran}

\begin{abstract}
\textbf{Abstract:} It is shown that the Bekenstein-Hawking entropy
of an axially symmetric dilaton-axion black hole can be expressed as
a Cardy-Verlinde formula. By utilizing the first order quantum
correction in the Bekenstein-Hawking entropy we find the modified
expressions for the Casimir energy and pure extensive energy. The
first order correction to the Cardy-Verlinde formula in the context
of axially symmetric dilaton-axion black hole are obtained with the
use of modified Casimir and pure extensive energies.
\end{abstract}
\maketitle

\newpage

\section{Introduction}

The entropy $S_{CFT}$ of conformal field theory (CFT) in an
arbitrary dimension $n$ has been related to its total energy $E$ and
Casimir energy $E_C$ by a relation, named as the Cardy-Verlinde
formula $S_{CFT}=\frac{2\pi R}{n}\sqrt{E_C(2E-E_C)}$
~\cite{verlinde}. The entropy associated with the conformal field
theory has been related to the Bekenstein-Hawking entropy for
various  black hole geometries with asymptotically anti-de Sitter
(AdS) boundary \cite{set1,set2,tab,kle,jamil,set3,set4,a1,jamil11}.
Thus, one may naively expect that the entropy of all CFTs that have
an AdS-dual description is given as the Cardy-Verlinde formula .
However, AdS black holes do not always satisfy the Cardy-Verlinde
formula \cite{gib}. Recently, much interest has been developed in
calculating the quantum corrections to the Bekenstein-Hawking
entropy $S$
 by using various techniques like radial null
geodesics, Hamilton-Jacobi method and loop quantum gravity etc
\cite{{m1},{m2},{m3}}. The leading-order correction is proportional
to $\ln S$ which comes out to be the same with the use of above
techniques. The leading order quantum correction to the classical
Cardy-Verlinde formula has been studied by Carlip \cite{carli}.

The thermodynamics of conformal field theories with gravity duals
has been studied actively in literature with the remarkable
resemblance of the relevant thermodynamic formulas
\cite{verlinde,set1,set2,tab,kle,jamil,set3,set4,a1,jamil11}. It has
been shown that the Cardy-Verlinde formula holds with a negative
cosmological constant or a more general certain potential term for
super-gravity scalars \cite{klemm}. There it has been argued that
the Cardy-Verlinde formula also holds for black hole geometry which
are asymptotically flat instead of asymptotically AdS space. In the
spirit of this Ref. \cite{klemm}, we discuss the entropy of
dilaton-axion black hole which is asymptotically flat spacetime in
terms of the Cardy-Verlinde formula. Here we consider the stationary
axially-symmetric axion-dilaton black hole to study the
Cardy-Verlinde formula and its first order correction. This black
hole is a string theory inspired black hole in lower spacetime
dimensions \cite{gar,jing}. The string theory inspired-models
consist of two massless scalar fields namely dilaton and axion, in
the low energy effective action in four dimension. The
thermodynamics of axially-symmetric axion-dilaton black hole is
investigated by various authors \cite{xxx}. We shall demonstrate
that the Cardy-Verlinde formula can be related with the
Bekenstein-Hawking entropy of the stationary axially-symmetric
axion-dilaton black hole. By employing the first order entropy
correction to Bekenstein-Hawking entropy, we are able to find the
leading order term of the Cardy-Verlinde formula.

The plan of the paper is: In the second section, we shall briefly
discuss the thermodynamic quantities associated with the horizon of
the stationary axially-symmetric dilaton-axion black hole. In third
section, we will study the entropy of the axially-symmetric
axion-dilaton black hole which can be represented by the
Cardy-Verlinde formula. In the fourth section, we provide the
leading order correction to the Cardy-Verlinde formula by using
quantum corrected Bekenstein-Hawking entropy in the context of
dilaton-axion black hole. Finally we shall conclude our results.

\section{Axially symmetric Einstein-Maxwell dilaton-axion black
hole}

In this section we shall consider the effective Lagrangian of the
low-energy heterotic string theory in four dimensions given by
\cite{garcia,jing}
\begin{eqnarray}
I&=&\frac{1}{16\pi}\int d^4x\sqrt{-g}\Big(
R-2g^{\mu\nu}\nabla_\mu\Phi\nabla_\nu\Phi-\frac{1}{2}e^{4\Phi}
g^{\mu\nu}\nabla_\mu K_a\nabla_\nu K_a\nonumber
\\&\;& -e^{-2\Phi} g^{\mu\lambda}g^{\nu\rho}F_{\mu\nu}F_{\lambda\rho}-
K_a F_{\mu\nu}\bar {F}^{\mu\nu}  \Big),
\end{eqnarray}
where the dual of electromagnetic field tensor $F_{\mu\nu}$ is
\begin{equation}
\bar {F}^{\mu\nu}
=-\frac{1}{2}\sqrt{-g}\varepsilon_{\mu\nu\alpha\beta}F^{\alpha\beta}.
\end{equation}
Here $R$ is the Riemann curvature scalar,
$\varepsilon_{\mu\nu\alpha\beta}$ is the Levi Civita symbol and
$g^{\mu\nu}$ is the metric tensor. Also $\Phi$ and $K_a$ are the
massless dilaton field and the axion field respectively.

In the Boyer-Lindquist coordinates $(t, r, \theta, \varphi)$, the
stationary axially-symmetric solution to the Einstein-Maxwell's
equations in the presence of the dilaton-axion is given by
\cite{jing},
\begin{eqnarray}\nonumber
ds^{2}&=& - \frac{\Sigma - a^{2}\sin^{2}\theta}{\Delta} dt^{2} -
\frac{2a \sin^{2}\theta}{\Delta}\left[(r^{2}
-2Dr+a^{2})-\Sigma\right]dt d\varphi\\ &&+ \frac{\Delta}{\Sigma}
dr^{2} + \Delta d\theta^{2}  + \frac{\sin^{2}\theta}{\Delta}
\left[(r^{2}-2Dr+a^{2})^2-\Sigma
a^{2}\sin^{2}\theta\right]d\varphi^{2},
\end{eqnarray}
where
\begin{equation}
\Delta = r^{2} -2Dr + a^{2}\cos^{2}\theta,\ \ \ \ \Sigma = r^{2}-2mr
+ a^{2},
\end{equation}
and
\begin{equation}
e^{2\Phi}=\frac{W}{\Delta}=\frac{\omega}{\Delta}(r^2+a^2\cos^2\theta),\
\ \ \omega=e^{2\Phi_0},
\end{equation}
\begin{equation}
K_a=K_0+\frac{2aD\cos\theta}{W},
\end{equation}
\begin{equation}
A_t=\frac{1}{\Delta}(Qr-ga\cos\theta),\ \ A_r=A_\theta=0,
\end{equation}
\begin{equation}
A_\varphi=\frac{1}{a\Delta}(-Qra^2\sin^2\theta+g(r^2+a^2)a\cos\theta).
\end{equation}
The mass $M$, angular momentum $J$, electric charge $Q$, and
magnetic charge $P$, dilaton charge $D$ of the black hole are given
by
\begin{equation}
M=m-D,\ \ J=a(m-D),\ \ Q=\sqrt{2\omega D(D-m)},\ \ P=g.
\end{equation}
The above results show that the stationary axis symmetric
dilaton-axion black hole significantly differs from the the
Kerr-Newmann black hole. The two horizons are the inner $r_-$ and
the outer one $r_+$ of the black hole under consideration are
\begin{equation}
r_\pm=M+D\pm\sqrt{(M+D)^2-a^2}.
\end{equation}
Only $r_+$ is the event horizon and one can associate
thermodynamical quantities with it.

The Hawking temperature associated with the event horizon is
\begin{eqnarray}
T& =& \frac{\hbar}{4\pi }\left(\frac{r_{+}-M-D}{r_{+}^{2} -2Dr_{+}+
a^{2}}\right).
\end{eqnarray}
The angular velocity $\Omega$ at the event horizon can be rewritten
as
\begin{eqnarray}
\Omega &=&\frac{J/M}{ r_{+}^{2} -2Dr_{+}+a^{2}}.
\end{eqnarray}
Here $J$ is the angular momentum. The electrostatic potential can be
given by
\begin{equation}
\Phi = \frac{-2DM}{Q(r_{+}^{2} -2Dr_{+}+ a^{2})}.
\end{equation}
The entropy associated with the event horizon of the dilaton-axion
black hole is
\begin{eqnarray}
S&=&\frac{\pi}{\hbar} (r_+^2-2Dr_++a^2).
\end{eqnarray}

\section{Cardy-Verlinde formula and dilaton-axion black hole}
In this section, we introduce the Cardy-Verlinde formula which
states that the entropy of a (1+1)-dimensional CFT is given by
\begin{equation}
S=2\pi\sqrt{\frac{c}{6}\Big( L_0-\frac{c}{24}  \Big)},
\end{equation}
where $c$ is the central charge and $L_0$ is the Virasoro generator.
After appropriate identifications of $c$ and $L_0$, the above Cardy
formula, we obtain the generalized Cardy-Verlinde formula which
takes the form \cite{verlinde}
\begin{equation}
S_{CFT}=\frac{2\pi R}{\sqrt{a_1b_1}}\sqrt{E_C(2E-E_C)},
\end{equation}
where $E$ is the total energy, $E_C$ is the Casimir energy, $a_1$
and $b_1$ are arbitrary positive constants. Also $R$ is the radius
of the $n+1$ dimensional spacetime, $ds^2=-dt^2+R^2d\Omega_n$.
 The definition of
Casimir energy is derived by the violation of the Euler relation as
\begin{equation}
E_C=n(E+PV-TS-\Phi Q-J\Omega),
\end{equation}
where the pressure of the CFT is given by $P=E/nV$. The total energy
is the sum of two terms
\begin{equation}
E(S,V)= E_E(S,V)+\frac{1}{2} E_C(S,V).
\end{equation}
Here $E_E$ is the purely extensive part of the total energy. The
Casimir energy and the purely extensive part of the total energy are
expressed as
\begin{equation}
E_C=\frac{b_1}{2\pi R}S^{1-\frac{1}{n}},
\end{equation}
\begin{equation}
E_E=\frac{a_1}{4\pi R}S^{1+\frac{1}{n}}.
\end{equation}

\section{Entropy of axially symmetric axion-dilaton black hole and Cardy-Verlinde formula}

Using Eq. (12) with $n=2$ and $E=M$, we obtain
\begin{eqnarray}
E_C&=&3M-2TS-2\Phi Q-2\Omega J,\nonumber\\
&=&3M-\frac{1}{2}(r_+-M-D)+\frac{4DM}{r_+^2-2Dr_++a^2}-\frac{2J^2}{M(r_+^2-2Dr_++a^2)}.
\end{eqnarray}
From (16) we have
\begin{eqnarray}
2E-E_C&=&-M+2TS+2\Phi Q+2\Omega J,\nonumber\\
&=&
-M+\frac{1}{2}(r_+-M-D)-\frac{4DM}{r_+^2-2Dr_++a^2}+\frac{2J^2}{M(r_+^2-2Dr_++a^2)}.
\end{eqnarray}
From (13) and (16), the extensive part of total energy becomes
\begin{eqnarray}
E_E&=&E-\frac{1}{2}E_C,\nonumber\\
&=&-\frac{1}{2}M+\frac{1}{4}(r_+-M-D)-\frac{2DM}{r_+^2-2Dr_++a^2}+\frac{J^2}{M(r_+^2-2Dr_++a^2)}.
\end{eqnarray}
Comparison of (14) and (16) yields
\begin{eqnarray}
R&=&\frac{b_1S^{1/2}}{2\pi}\Big[3M-2TS-2\Phi Q-2\Omega J\Big]^{-1},\nonumber\\
&=&\frac{\frac{b_1}{2\pi}\sqrt{\frac{\pi}{\hbar}
(r_+^2-2Dr_++a^2)}}{3M-\frac{1}{2}(r_+-M-D)+\frac{4DM}{r_+^2-2Dr_++a^2}-\frac{2J^2}{M(r_+^2-2Dr_++a^2)}}.
\end{eqnarray}
Comparison of (15) and (18) yields
\begin{eqnarray}
R&=&\frac{a_1S^{3/2}}{4\pi}\Big[-\frac{1}{2}M+TS+\Phi Q+\Omega J\Big]^{-1},\nonumber\\
&=&\frac{\frac{a_1}{4\pi}\Big[\frac{\pi}{\hbar}
(r_+^2-2Dr_++a^2)\Big]^{3/2}}{-\frac{1}{2}M+\frac{1}{4}(r_+-M-D)-\frac{2DM}{r_+^2-2Dr_++a^2}+\frac{J^2}{M(r_+^2-2Dr_++a^2)}}.
\end{eqnarray}
Combining the last two expressions (19) and (20), we obtain
\begin{eqnarray}
R&=&\frac{\sqrt{a_1b_1}}{2\sqrt{2}}\frac{\pi}{\hbar}
(r_+^2-2Dr_++a^2)\Big[
3M-\frac{1}{2}(r_+-M-D)+\frac{4DM}{r_+^2-2Dr_++a^2}-\frac{2J^2}{M(r_+^2-2Dr_++a^2)}
\Big]^{-1}\nonumber\\&&\times\Big[
-\frac{1}{2}M+\frac{1}{4}(r_+-M-D)-\frac{2DM}{r_+^2-2Dr_++a^2}+\frac{J^2}{M(r_+^2-2Dr_++a^2)}
\Big]^{-1}.
\end{eqnarray}
Using (16), (17) and (21) in (11) yields
\begin{equation}
S_{CFT}=\frac{\pi}{\hbar} (r_+^2-2Dr_++a^2)=S.
\end{equation}

\section{Logarithmic correction to the Cardy-Verlinde formula}

In this section, we shall obtain the first order entropy correction
by using corrected Bekenstein-Hawking entropy formula in the
Cardy-Verlinde formula. The first order correction to the
semi-classical Bekenstein-Hawking entropy $S_0$ is given by
\cite{setare}
\begin{equation}
\mathbb{S}=S_0-\frac{1}{2}\ln C.
\end{equation}
Here $C$ is the heat capacity of the black hole evaluated at the
event horizon. We suppose that $C\simeq S=S_0$ \cite{setare} so that
the above equation (28) turns out
\begin{equation}
\mathbb{S}=S_0-\frac{1}{2}\ln S_0.
\end{equation}
First we calculate the corrected Casimir energy and the corrected
extensive part of the total energy by using first order corrected
entropy (29) which admit
\begin{equation}
\tilde{E}_C=E_C+T\ln S_0,
\end{equation}
\begin{equation}
\tilde{E}_E=E-\frac{1}{2}E_C-\frac{1}{2}T\ln S_0.
\end{equation}
By using modified Casimir energy (30) and the extensive part of the
total energy (31) in the Cardy-Verlinde formula (16), we obtain the
modified Cardy-Verlinde entropy relation
\begin{equation}
\tilde{S}_0=\frac{2\pi
R}{\sqrt{a_1b_1}}\sqrt{\tilde{E}_C(2E-\tilde{E}_C)}.
\end{equation}
Simplifying (32) we obtain
\begin{equation}
\tilde{S}_0\simeq S_0\Big[ 1+\frac{(E-E_C)}{E_C(2E-E_C)} T\ln S_0
\Big].
\end{equation}
Finally using (33) in (29) yields the corrected entropy as
\begin{equation}
\mathbb{S}\simeq\frac{2\pi
R}{\sqrt{a_1b_1}}\sqrt{E_C(2E-E_C)}+\Big[\frac{2\pi
R}{\sqrt{a_1b_1}}\frac{(E-E_C)}{\sqrt{E_C(2E-E_C)}}-\frac{1}{2}\Big]T\ln\Big[\frac{2\pi
R}{\sqrt{a_1b_1}}\sqrt{E_C(2E-E_C)} \Big].
\end{equation}
Hence the entropy correction to the semi-classical
Bekenstein-Hawking entropy is obtained in terms of the modified
Cardy-Verlinde formula which further investigates the AdS/CFT
correspondence in terms of modified Cardy-Verlinde entropy formula.
The first term corresponds to the usual CV formula while the second
term relates to correction to Hawking entropy in terms of modified
Cardy-Verlinde entropy formula.

\section{Conclusion}

In this paper, we have shown that the Bekenstein-Hawking entropy of
the axially-symmetric axion-dilaton black hole can also be expressed
in the form of Cardy-Verlinde entropy formula which further
investigates the AdS/CFT correspondence in terms of Cardy-Verlinde
entropy formula. The axially symmetric dilaton axion black hole is
asymptotically flat instead of AdS space. So our study indicates
that the AdS/CFT correspondence still holds in the black hole
geometries with asymptotically flat background. By using the
logarithmic correction to the Bekenstein-Hawking entropy, we
obtained the modified expressions for the Casimir and extensive
energy relations. By utilizing modified expressions for Casimir and
extensive energy in the Cardy-Verlinde formula, we obtained the
corrected $S_{CFT}$ relation which relates the entropy of a certain
CFT to its total energy and Casimir energy. The second result of
this paper is the entropy correction to the semi-classical
Bekenstein-Hawking entropy in terms of the modified Cardy-Verlinde
formula. The first term in (34) corresponds to the usual
Cardy-Verlinde formula while the second term relates correction to
Hawking entropy in terms of modified Cardy-Verlinde entropy formula.


\begin{thebibliography}{99}

\bibitem{verlinde} E. Verlinde, hep-th/0008140;\\ M.R. Setare and
E.C. Vagenas, Phys. Rev. D 68 (2003) 064014;\\ R-G Cai, Phys. Lett.
B 525 (2002) 331;\\ R-G. Cai, Nucl. Phys. B 628 (2002) 375.
\bibitem{set1} M.R. Setare, Mod. Phys. Lett. A 17 (2002) 2089.
\bibitem{set2} M.R. Setare, and M.B. Altaie, Eur. Phys. J. C
               30 (2003) 273.
\bibitem{tab} D. Birmingham and S. Mokhtari, Phys. Lett. B 508
              (2001) 365;\\ C.O. Lee, Phys. Lett. B 670 (2008) 146.
\bibitem{kle} D. Klemm et al, Nucl. Phys. B 601 (2001) 380.
\bibitem{jamil} M.R. Setare and M. Jamil, Phys. Lett. B 681 (2009)
471.
\bibitem{set3} M.R. Setare and R. Mansouri, Int. J. Mod. Phys. A 18
 (2003) 4443.
\bibitem{set4} M.R. Setare and E.C. Vagenas, Int. J. Mod. Phys. A 20
(2005) 7219.
\bibitem{a1} B. Wang et al, Phys. Lett. B 503
 (2001) 394.
\bibitem{jamil11} M.R. Setare and M. Jamil, arXiv:1001.4716
\bibitem{gib} G.W. Gibbons et al, Phys. Rev. D 72 (2005) 084028.
\bibitem{m1}A.J.M. Medved, Class. Quant. Grav. 19 (2002) 2503.
\bibitem{m2}S. Mukherji and S.S. Pal, JHEP 0205 (2002) 026.
\bibitem{m3}J.E. Lidsey et al, Phys. Lett. B 544 (2002)
337.
\bibitem{carli}S. Carlip, Class. Quant. Grav. 17 (2000) 4175.
\bibitem{klemm} D. Klemm et al, hep-th/0104141
\bibitem{gar} D. Garfinkle et al, Phys. Rev. D 43 (1991) 3140.
\bibitem{jing} J. Jing, Nuc. Phys. B 476 (1996) 548.
\bibitem{xxx}  G.A.S. Dias and J.P.S. Lemos,
Phys. Rev. D 78 (2008) 084020;\\ Y.S. Myung et al, Phys. Lett. B 663
 (2008) 342;\\ A. Sheykhi, Phys. Rev. D 76 (2007) 124025;\\ G. Kunstatter et al,
  Phys.Rev. D 57 (1998) 3537.
\bibitem{garcia} A. Garcia et al, Phys. Rev. Lett. 74 (1995) 1276
\bibitem{setare}R. K. Kaul and P. Majumdar, Phys. Rev. Lett. 84 (2000)
5255;\\
S. Das, P. Majumdar and R.K. Bhaduri, Class. Quant. Grav. 19 (2002)
2355;\\ M. R. Setare, Eur. Phys. J. C 33 (2004) 555.
\end{thebibliography}
\end{document}